\documentclass[aps,prl,twocolumn,floatfix]{revtex4}
\usepackage{amssymb}

\usepackage{graphicx}

\begin{document}

\title{Convective dust clouds in a complex plasma}
\author{S. Mitic, R. S\"{u}tterlin, A.V.Ivlev, H. H\"{o}fner, M.H. Thoma, S. Zhdanov, G. E. Morfill}
%\email{mitic@mpe.mpg.de}
\affiliation{Max-Planck-Institut f\"ur
extraterrestrische Physik, D-85741 Garching, Germany}

\date{\today}

\begin{abstract}
The plasma is generated in a low frequency glow discharge within an
elongated glass tube oriented vertically. The dust particles added
to the plasma are confined above the heater and form
counter-rotating clouds close to the tube centre. The shape of the
clouds and the velocity field of the conveying dust particles are
determined. The forces acting on the particles are calculated. It is
shown that convection of the dust is affected by the convective gas
motion which is triggered, in turn, by thermal creep of the gas
along the inhomogeneously heated walls of the tube.

\end{abstract}

\maketitle

%\section{Introduction}

Convective motion of micro particles in complex (dusty) plasmas is
a phenomenon that is often observed in very different experimental
conditions. In particular, vortices in complex plasmas can be
produced both in ground-based laboratories and under microgravity
conditions, in dc and rf discharges of fairly different
configuration, upon inhomogeneous heating and in rather isothermal
environment \cite{Fortov05}.

There are many publications, both theoretical and
experimental, in which the origin of the vortex motion in complex
plasmas has been investigated \cite{Vaulina00,Shukla00,
Tsytovich06,Antipov06,Zuzic07,Fortov03,Vaulina03,Morfill99}.
Basically, there are two mechanisms that can produce vortices:
This is either the presence of non-potential force(s) exerted on
charged micro particles in the discharge (due to inhomogeneous
charges \cite{Fortov03} or because of ion drag \cite{Morfill99})
or the convective motion of the background neutral gas or dust
particles themselves \cite{Ivlev07}. This clearly indicates that
the nature of such vortices -- despite of quite similar appearance
-- might be very different as well.

In this paper we report on a recent experiment performed in a low
frequency glow discharge under gravity. The PK-4 setup
\cite{Fortov2005, Khrapak2005} was used, which represents a
typical configuration of a dc discharge employed to study complex
plasmas. We investigate convective motion of micro particles in the
presence of an external controllable heating and unambiguously
show that under such conditions vortices in complex plasmas occur due
to the neutral gas convection. Clouds of micro particles in this
case resemble convective clouds in the atmosphere produced from warm
air pockets rising upwards and composed of water droplets, ice
crystals, ice pellets, etc.

%\section{Experiment}

\begin{figure}[tbp]
\centering
\includegraphics[width=2.0in]{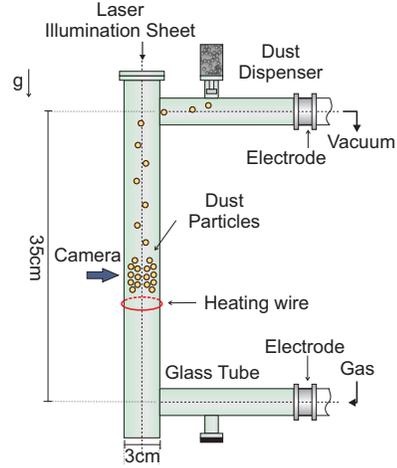}
\caption{Experimental setup.}
\label{fig:sketch}
\end{figure}

The experiments were conducted in a complex plasma produced by a
low frequency discharge (LFD) in Neon gas of the PK-4 facility \cite{Fortov2005,
Khrapak2005}. The plasma chamber consists of an elongated glass tube
as shown in figure~\ref{fig:sketch}.
To remove all rest gases which might disturb the experiments the
chamber was evacuated for several hours to reach a base pressure
below $10^{-3} Pa$. Neon gas with constant pressure between $30 Pa$
and $100 Pa$ were used during the experiments.

A gas discharge was maintained between the electrodes by a regulated discharge current of $1 mA$. The discharge voltage
polarity was changed with a frequency of $1 kHz$ ($50\%$ duty cycle). This frequency is more than an order of magnitude
higher than the dust-plasma (response) frequency \cite{Fortov05}, so that the effective longitudinal force on micro
particles (electric plus ion drag) vanishes.

In course of the experiments micro particles (spherical Melamine-Formaldehyde particles with a
mass density of $1.51 {g}/{cm^3}$) of two different radii $r_p=1.64 \mu m$ or $3.05 \mu m$  were injected by a dispenser from above
into the plasma and confined inside of the vertical glass tube. The
micro particles were illuminated along the tube axis by a laser sheet  (wave length
$686nm$, power $20 mW$, thickness $100 \mu m \pm 30 \mu m$, width $15 mm$)
and the images were recorded by a PCO 1600 camera.

A metal wire (alloy of $70\%$ Ni, $11\%$ Fe, and $14\%$ Cu and $0.5
mm$ in diameter) was put around the lower part of the tube to
produce a temperature gradient. A dc current between $0.5A$ and $1.8
A$ was applied to this wire using a $10V$ power supply. For avoiding
unwanted electromagnetic effects on the plasma only one loop of the
wire was attached. The wire was insulated against direct contact to
the tube walls. The resulting temperature distribution was
simultaneously measured at eight points on the glass tube at and
above the heating wire to extract the temperature gradient over a
range of $8 cm$.

Due to the gas temperature gradient a thermophoretic force acts on
the micro particles~\cite{Waldmann1959}. This was used to levitate
micro particles against gravity in complex plasma experiments
~\cite{Rothermel2002}. The thermophoretic force acting on the
micro particles is given by
\begin{equation}
{\bf F}_{th}=-\gamma_{th} {\nabla}T_n, \label{thermophoresis}
\end{equation}
where $\gamma_{th}[{\rm N\cdot cm/K}]\simeq2.17\times10^{-6}\, (r_p[\mu{\rm m}])^2$ ~\cite{Rothermel2002}. The
thermophoretic force required to levitate particles is $F_{th}=m_pg\simeq2.74\times 10^{-13} N$ for the smaller ones and
$\simeq1.76\times 10^{-12} N$ for the larger ones ($m_p$ is the particle mass). Therefore according to
(\ref{thermophoresis}) temperature gradients of $|dT/dz|\simeq4.65 {K}/{cm}$ for the smaller particles and $\simeq8.64
{K}/{cm}$ for the larger ones are required to compensate gravity.

\begin{figure}[tbp]
 \centering
 \includegraphics [width=2.5in]{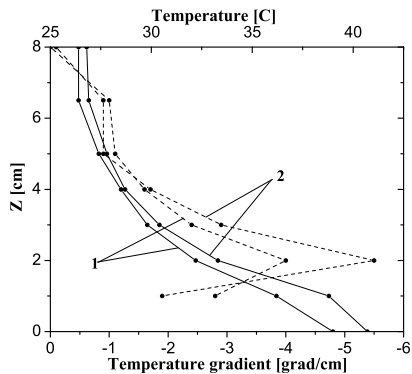}
 \caption{Measured temperatures (solid lines) and calculated temperature
   gradients (dashed lines).  1: temperature and temperature gradient for $r_p=1.64 \mu m$ particles, 2: same for $r_p=3.05 \mu m$ particles.}

%    The left curves of the temperature (solid lines) and the
%temperature gradient (dashed lines) correspond to experiments with
%particles of radius $r_p=1.64 \mu m$, the right ones to the $3.05
%\mu m$ particles.?????????correct fig and caption.}
 \label{fig:tgrad}
\end{figure}

\begin{figure*}[tbp]
 \includegraphics[width=1.5in]{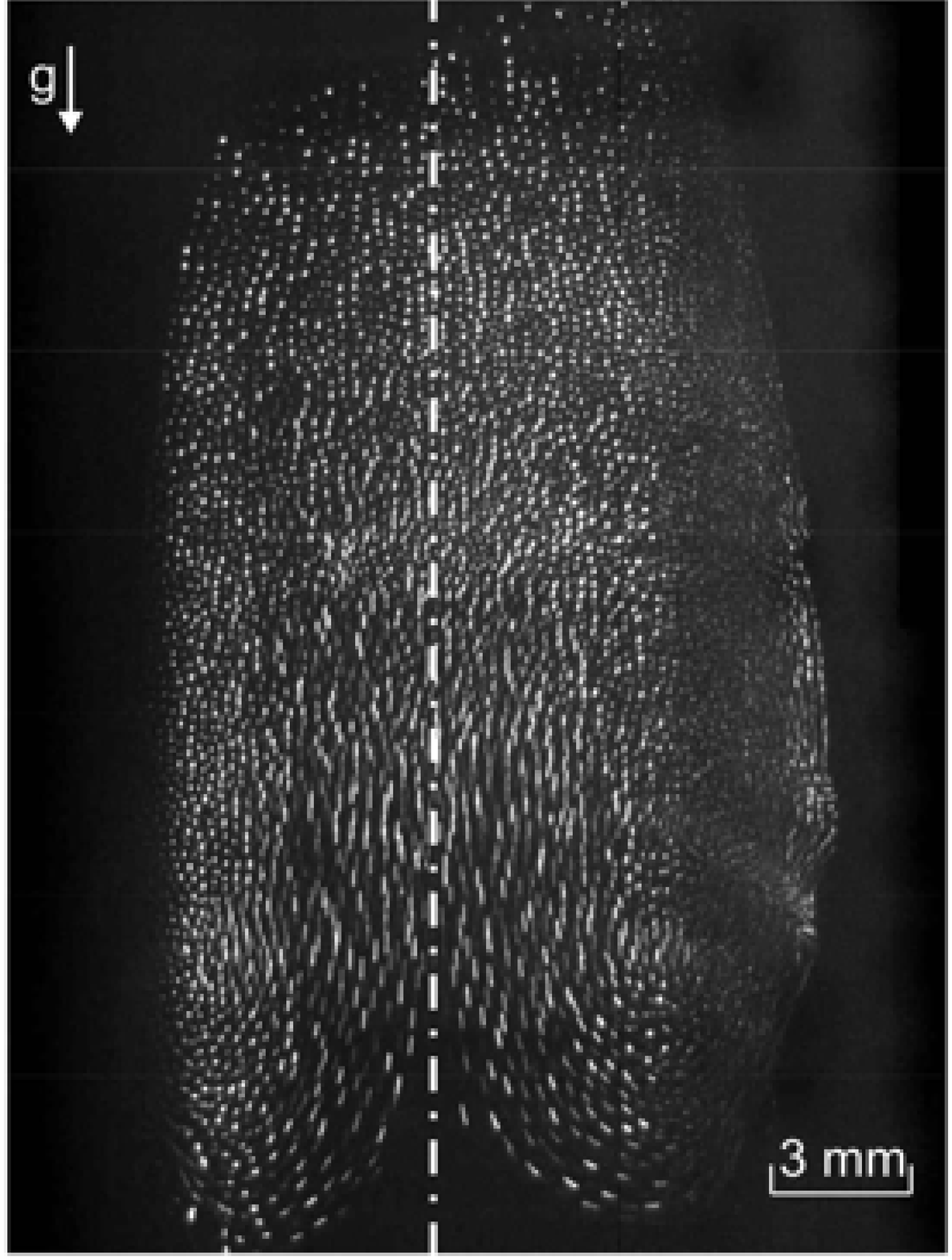}
 \includegraphics[width=1.5in]{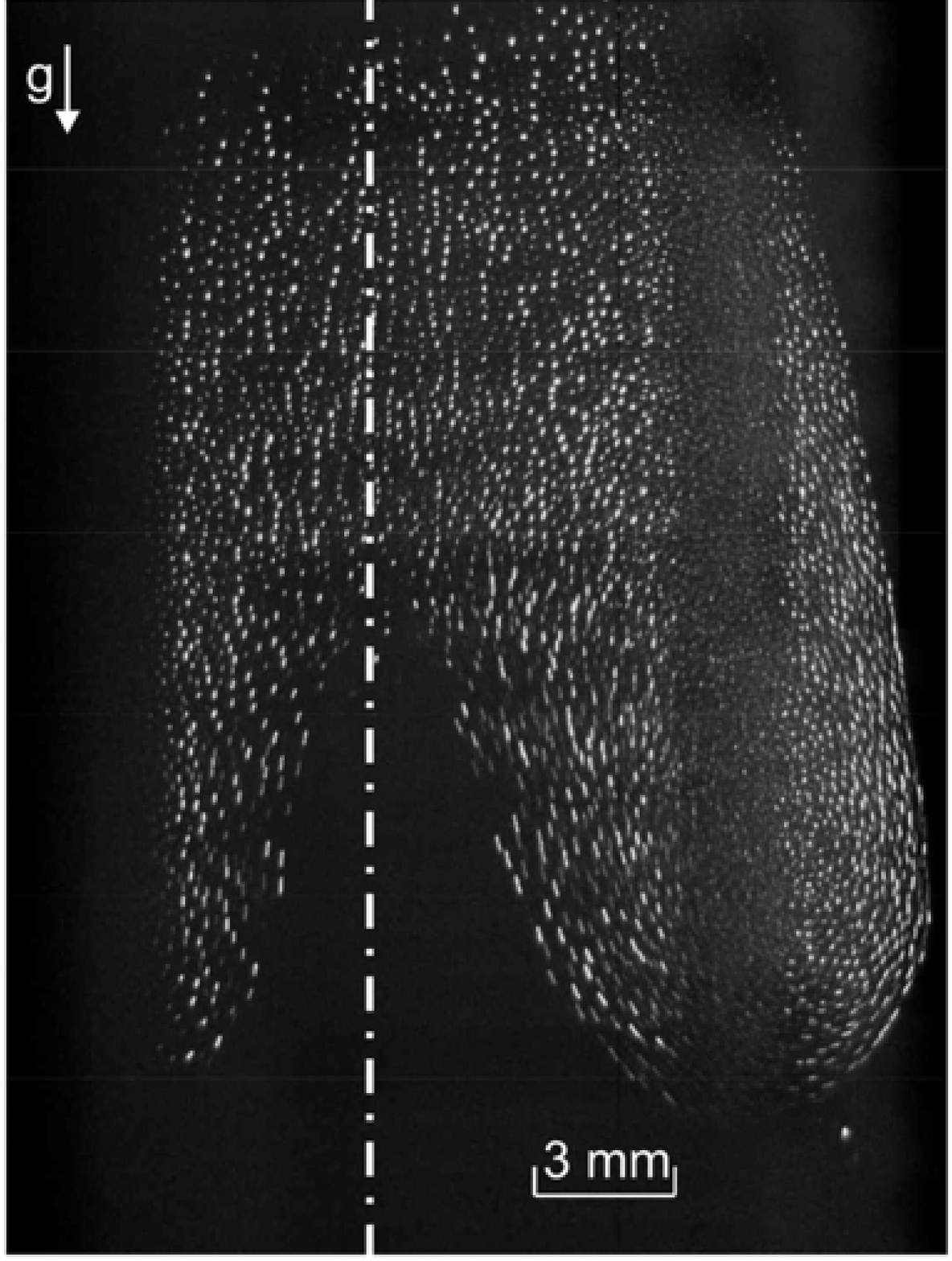}
 \includegraphics[width=1.5in]{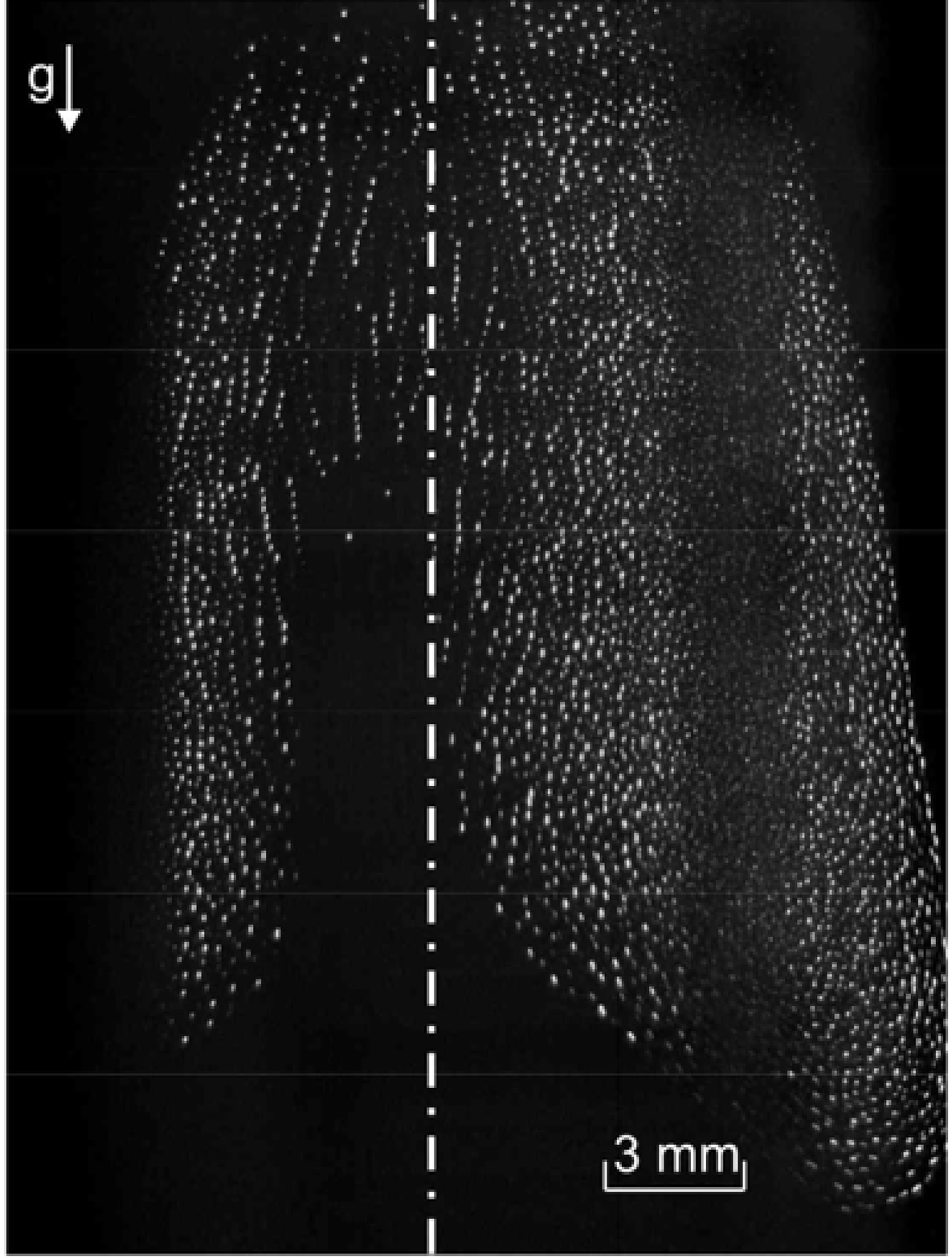}
 \caption{Dust clouds of $r_p=1.64 \mu m$ particles at $30$, $50$, and
   $100 Pa$ (from left to right). The dash-dotted lines indicate the center of the tube.
%Note that the central void in the bottom of the cloud becomes bigger with pressure increase
%indicating intensification of vertical gas flow.
The field of view is 21 x 26 mm. }
 \label{fig:convection}
\end{figure*}

The measured temperatures along the tube and the temperature gradients are shown in figure~\ref{fig:tgrad}, where $z=0$ is
the position of the heating wire. It turns out that the measured maximum values of the temperature gradients are smaller
than those necessary for thermophoretic levitation. Nevertheless we observe that the particle cloud is confined in the area
above the heating wire. Figure~\ref{fig:convection} shows vertical cross sections through the center of the clouds obtained
for different pressures. As we pointed out above, plasma forces cannot be responsible for this effect: because of the fast
polarity switching, neither a longitudinal electric force nor an ion drag force \cite{Ivlev04} could support the levitation.
In addition, an intense convective motion of the particles is observed within the levitated clouds.

The micro particles are frictionally coupled to the neutral gas. Therefore, it is natural to assume that gas convection is induced
by the local tube heating and that the micro particles are dragged by the gas motion. Favorably directed, the gas flow could
help to levitate the particles in addition to thermophoresis and also could result in global dust rotation. Note that ${\bf
F}_{th}$ alone cannot cause the convection, because it is a potential force, ${\nabla}\times{\bf F}_{th}\equiv0$.

%This assumption allows us to extract the velocity distribution of the convection as we will discuss in the following.

The neutral drag force acting on a spherical particle within a gas
flow is given by~\cite{Epstein1924}
\begin{equation}
{\bf F}_n=\gamma_n ({\bf v}_f - {\bf v}_p), \label{drag}
\end{equation}
where the Epstein friction coefficient reads
$
%\begin{equation}
\gamma_n[{\rm N\cdot s/cm}]\simeq2.7\times 10^{-16}\, (r_p[\mu{\rm m}])^2\, p[{\rm Pa}]\label{friction} $
%\end{equation}
for Neon at room temperature, $T_n =300 K$.

The radial electric field of discharge exerts an additional force $F_r$ \cite{Raizer} which confines the particle cloud in
radial direction. We observed that the micro particles after switching off the discharge expand across the tube, exploiting
practically the entire convective field. Particles keep rotating one or two cycles before eventually falling down.

%\section{Analysis}

As an example, let us extract the velocity field ${\bf v}_f$ of the convective gas flow and the radial electric force ${\bf
F}_r$ for the experiment with $3.05\mu m$ particles at a pressure of $50 Pa$.

Particle motion is recorded at a frame rate of $500 fps$ in course
of the experiment comprising two separate cases: with and without
plasma. The particles are detected in each frame and then based on
the position of each particle in a few consecutive frames their
velocities and accelerations are extracted. Based on this data, the
particle velocity profiles are reconstructed in the entire
convective dust cloud. Particle velocities and accelerations are
derived by fitting cubic splines to the complete particle
trajectories.

\begin{figure*}[tbp]
 \centering
 \includegraphics[width=5.0in]{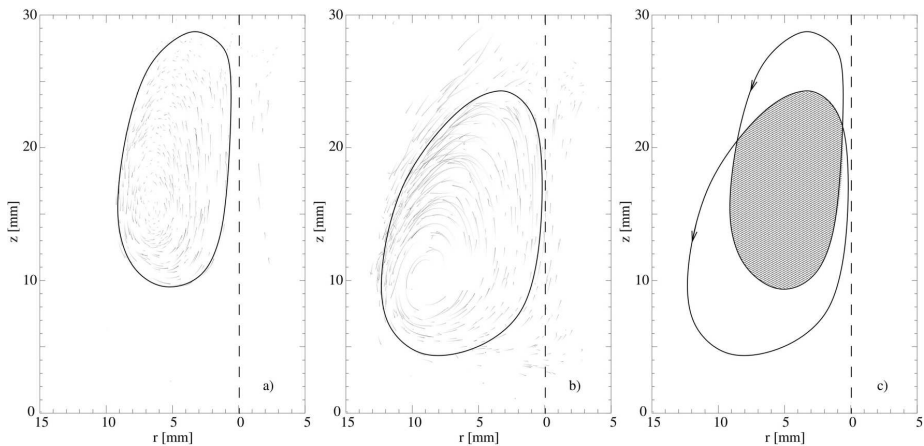}
 \caption{Convective dust clouds of $r_p=3.05\mu m$ particles at a pressure $50 Pa$
 for two different cases,
 (a) with switching on of the plasma
 and (b) with switching off the plasma.
 (c) The superimposed cloud positions for these two cases. The arrows indicate the directions of particle
 rotations. The region of overlapping trajectories is shaded. The dashed lines indicate
 the center of the tube.}
 \label{fig:trajectories}
\end{figure*}

The trajectories of individual particles are determined by
the balance of forces:
\begin{equation}
m_p{\ddot{\bf r}}={\bf F}_{th}+{\bf F}_n+{\bf F}_{r}+m_p {\bf g}. \label{force_equation}
\end{equation}
To separate the remaining unknown forces -- the radial electric force and the neutral drag force -- appearing in
(\ref{force_equation}), we analyze first the data when no plasma was in the tube, and, hence, no radial electric force
$(F_r=0)$ acts on the particles. This allows us to get the velocity field of the conveying gas molecules. Next with this
information and assuming that switching on and off the plasma does not alter the gas convection, we calculate the radial
electric force of the plasma in the region where particle trajectories recorded with and without plasma overlap (cf. figure
\ref{fig:trajectories}).
Switching off the plasma is a very fast process compared to the time particles spend conveying in the tube. Particles
trajectories for both cases are shown in figure \ref{fig:trajectories}. In 
figure \ref{fig:trajectories}(a) the
particle cloud in the presence of the plasma is a bit higher and further 
away from the tube wall. After switching off the plasma \ref{fig:trajectories}(b)
particles come closer to the wall and the cloud expands further down. The area of particle clouds and the direction of
rotation for these two cases are represented in (c) with the overlapping region marked. For the overlapping region any
differences in the particle motion between the cases with and without plasma must be due to electrical forces as the
distribution of temperature and free gas convection are not influenced by the discharge.

The neutral drag force on the particles depends
on the relative velocities of the particles and gas.
This is the reason why
the particles do not follow the gas flow trajectories but rotate eccentrically from the center of gas convection
in the area of upward gas draft.
\begin{figure}[tbp]
 \centering
 \includegraphics[width=2.0in]{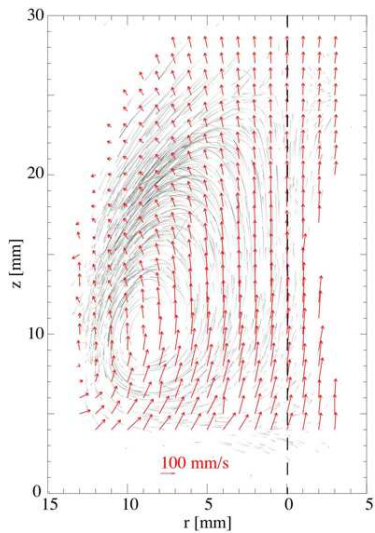}
 \caption{Averaged gas flow velocity field (vectors) superimposed with particles trajectories. The dashed line correspond to the center of the tube. }
 \label{fig:flow}
\end{figure}
Figure~\ref{fig:flow} presents the velocity field of the free gas convection
as calculated for the case without plasma.
The particle trajectories are overplotted. The distribution of flow velocities has a clear rotational tendency.

%\begin{figure}[tbp]
% \centering
%  \includegraphics[width=0.4\textwidth]{fig_xflow_ypos.eps}
%  \hspace{1cm}
%  \includegraphics[width=0.4\textwidth]{fig_xpos_yflow.eps}
%  \caption{Gas flow velocity axial distributions which is characteristic for rotational motion.}
%  \label{fig:flowpos}
%\end{figure}

\begin{figure}[tbp]
 \centering
 \includegraphics[width=2.0in]{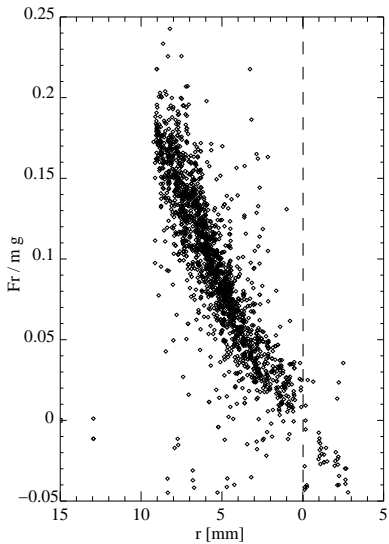}
 \caption{Reconstructed radial force for the overlapping region. The dashed line indicate the center of the tube.}
 \label{fig:radial}
\end{figure}

The radial force on the particles when plasma is on (cf. figure~\ref{fig:radial})
can now be
fitted using the gas flow velocities in the overlapping region.
The radial force is zero at the tube axis and rapidly increases
towards the tube walls.

%\section{Discussion}

The results obtained above clearly demonstrate the presence of gas convection. This is not free convection, though, merely
because the onset of free convection is too high in terms of the critical Reyleigh number (see, for instance
~\cite{Ivlev07}) to cause gas flow under the conditions of our experiments. It is well known, however, that if one puts a
non-uniformly heated body in a rarefied gas, the gas starts moving along the body {\it in the direction} of the temperature
gradient ~\cite{Kogan69, Bakanov92, Lifshits81}. This phenomenon is referred to as \emph{thermal gas creep} (or
\emph{thermal gas slip}). It was predicted theoretically by Maxwell ~\cite{Maxwell1879} and verified experimentally by
Reynolds ~\cite{Reynolds1880} and is governed by the relation
\begin{equation}
V_{TC} = K_{TC}\nu \nabla _{\|}\ln T_W, \label{creep}
\end{equation}
where $V_{TC}$ is the velocity of creep at the body surface, $\nu$
is the kinematic viscosity of the gas, $T_W$ is the temperature of
the heated body, i.e. the glass walls in our case,
and $\|$ indicates the component of the gradient along the
surface. For a long tube, the radial distribution of the (longitudinal)
velocities is well known,
%\begin{equation}
$v_{z} = V_{TC}(2r^2/R^2-1) \label{gas speed}$
%\end{equation}
~\cite{Lifshits81}. In our geometry the vertical temperature gradient is negative  $dT_W/dz<0$ (see figure \ref{fig:tgrad}).
Hence, the gas should flow downwards along the tube walls with $v_z(R)=V_{TC}$ and upwards near the tube axis with
$v_z(0)=-V_{TC}$,
%\begin{equation}
%v_{z} \equiv v_0=-V_{TC}>0, \label{gas axis speed}
%\end{equation}
which is in agreement with our observations (see figure \ref{fig:flow}). For the quantitative comparison with the
experiment, we rewrite Eq. (\ref{creep}) in the following form:
%\begin{equation}
$K_{TC}=|V_{TC}/\nu \nabla _{z}\ln T_W|$.
%\end{equation}
Based on the experimental data shown in figures \ref{fig:tgrad} and \ref{fig:flow} we get $K_{TC}\simeq 1$, which is in
agreement with theoretical expectations. (The theoretically possible range of values of $K_{TC}$ amounts to 0.7--1.2
~\cite{Bakanov92}). Note also that the magnitude of the convection velocity decreased monotonically with pressure, which is
also in line with the theoretical estimates: combining Eqs. (\ref{force_equation}) and (\ref{creep}), we obtain that the
velocity should scale as $\propto p^{-1}$.

%\section{Conclusions}

To conclude, we observed levitation of a particle cloud in a vertical glass tube, above a heated wire. In addition, the
particles exhibited a global vortex flow. We showed that the particle vortices were induced by the convection of neutral gas
analogously to convective clouds in the atmosphere. In turn, the gas convection was triggered by the thermal creep along
the inhomogeneously heated tube surface.

%\section{ACKNOWLEDGMENTS}
%{\bf ACKNOWLEDGMENTS}

The authors would like to acknowledge valuable discussions with Uwe Konopka.
This work was supported by DLR under grant 50 WM 0504.

\end{document}